# Infall in massive clumps harboring bright infrared sources

Ying-Hua Yue[1], Sheng-Li Qin[1], Tie Liu[2], Meng-Yao Tang[1], Yuefang Wu[3], Ke Wang[4], Chao Zhang[1]

[1] Department of Astronomy, Yunnan University, and Key Laboratory of Astroparticle Physics of Yunnan Province, Kunming, 650091, People's Republic of China *yue_yinghua@yeah.net*

[2] Shanghai Astronomical Observatory, Chinese Academy of Sciences, 80 Nandan Road, Shanghai 20030, People's republic of China

[3] Department of Astronomy, Peking University, 100871, Beijing, People's Republic of China

[4] Kavli Institute for Astronomy and Astrophysics, Peking University, 5 Yiheyuan Road, Haidian District, Beijing 100871, China

**Abstract** Thirty massive clumps associated with bright infrared sources were observed to detect the infall signatures and characterize infall properties in the envelope of the massive clumps by APEX telescope in CO(4-3) and $C^{17}O$(3-2) lines. Eighteen objects have "blue profile" in CO(4-3) line with virial parameters less than 2, suggesting that global collapse is taking place in these massive clumps. The CO(4-3) lines were fitted by the two-layer model in order to obtain infall velocities and mass infall rates. Derived mass infall rates are from $10^{-3}$ to $10^{-1}$ $M_\odot yr^{-1}$. A positive relationship between clump mass and infall rate appears to indicate that gravity plays a dominant role in the collapsing process. Higher luminosity clump has larger mass infall rate, implying that the clump with higher mass infall rate has higher star formation rate.

## 1 INTRODUCTION

Both observational and theoretical studies have suggested that low-mass stars form through collapse, outflow and disk accretion processes in molecular cloud cores (Shu et al. 1987). How massive stars form is not well understood mainly due to short lifetime, large distance and cluster environment. There are two promising models to explain the massive star formation mechanism (Tan et al. 2014). One is "core accretion", which suggests that isolated massive gas cores form high-mass stars with a larger accretion rate as do low-mass stars (McKee & Tan 2003). Another one is "competitive accretion", which claims that high-mass stars form in the center of a cluster and compete with off-center protostars for cloud gas (Bonnell et al. 2001). In any cases, observations of global collapse in massive clumps are critical for studying massive star formation, since most of massive stars are formed in clusters which are related to fragmentation of massive molecular clumps (Peretto et al. 2013)

The blue profile and red-shifted absorption line profile are generally considered as a key signature of infall (Evans 2003; Wyrowski et al. 2012). "Blue profile" says that optically thick line profile is double-peaked with a central self-absorption dip and the blue peak is stronger than the red one, while the optically thin line peaks in the self-absorption dip of the double-peaked line.

Infall is firstly identified in low-mass protostar (Zhou et al. 1993), and then frequently detected in low-mass star-forming regions. Up to now, lots of infall candidates are identified in massive clumps and cores by both single dish and interferometers by use of various tracers (Wu & Evans 2003; Fuller et al. 2005; He et al. 2015; Wu et al. 2007; Zapata et al. 2008; Qin et al. 2016; Wu et al. 2009; Liu et al. 2011; Qiu et al. 2012; Liu et al. 2013; Tang et al. 2019; Yang et al. 2020). The observations suggested that material around high-mass sources is infalling similar with low mass protostars, but high-mass sources have higher mass infall rates. However previous large sample single dish observations usually use lower energy transition lines which trace outer envelopes of massive clumps and cores. Interferometric observations are only made towards a few cases by employing different tracers which is difficult for comparison.

Massive stars form in densest part of the clumps or cores. High energy transitions of the molecules trace dense gas environments and are more likely to produce blue asymmetric emission profiles (Chira et al. 2014). Mid-J CO transitions with a stable abundance in warm regions is a better tracer of physical conditions in the inner envelope of massive clumps. The CO(4-3) observations clearly show that gas infall is ongoing in massive clumps in massive clumps associated with UC Hii regions (Wyrowski et al. 2006).

Previous observations towards the massive clumps at different evolutionary stages suggested the sources associated with UC Hii regions have higher detection rate of infall (Wu et al. 2007). In this paper, 30 massive molecular clumps harboring bright infrared sources with different mass and luminosity are observed with APEX telescopes in CO(4-3) and $C^{17}O(3-2)$ lines. The infrared color of the 30 sources is characteristic of UC Hii region, and the 30 sources are thought to be candidates of UC Hii region (Faúndez et al. 2004). The main goals are to investigate infall properties in the envelope of the massive clumps associated with bright infrared sources and the relationship among mass infall rate and mass, luminosity.

## 2 OBSERVATIONS

The observations were performed with the Atacama Pathfinder Experiment (Güsten et al. 2006, APEX) in CO(4-3) and $C^{17}O(3-2)$ lines from 8 to 10 August, 2015. The antenna system temperatures range from 236 to 255 K, and from 865 to 1145 K respectively, during $C^{17}O(3-2)$ and CO(4-3) observations. The frontend used was the MPIFR dual channel FLASH receiver (460 and 810 GHz) and APEX2 receiver (345 GHz) (Heyminck et al. 2006), along with the APEX digital Fast Fourier Transform Spectrometer backends (Klein et al. 2006, FFTS) was used for lines observations. The lines were covered with 2048 channels within a bandwidth of 1 GHz. For a better view, we have smoothed the spectral resolution to 0.27 and 0.4 km s$^{-1}$ for CO (4-3) and $C^{17}O(3-2)$ lines.The correction for the line intensities to main beam temperature scale was made using the formula $T_{mb} = (F_{eff}/B_{eff})T_A$. The forward efficiencies ($F_{eff}$) of 0.95 and beam efficiencies of 0.73 and 0.6 at 345 and 460 GHz were used, respectively. Mean 1 $\sigma$ rms noise levels are 0.19 and 0.14 K for CO(4-3) and $C^{17}O(3-2)$, respectively. The beam sizes are 13″ at 461 GHz(CO(4-3)) and 18″ at 337 GHz ($C^{17}O(3-2)$). All observations were done in position switching model. The 30 sources were selected from Atacama Submillimeter Telescope Experiment (ASTE) observational sample with blue profile in HCN line (Liu et al. 2016). The source name, coordinates, bolometric luminosity (L), masses (M), volume density (n), effective radii (R) are summarized in Table 1.

## 3 RESULTS

Table 1: Physical parameters of the observed sources.

| Source Name | RA | DEC | L $L_\odot$ | M $M_\odot$ | $n_{H_2}$ cm$^{-3}$ | R pc | $V_{C^{17}O}$ km s$^{-1}$ | $\Delta V_{C^{17}O}$ km s$^{-1}$ | $\delta v$ | $V_{in}$ km s$^{-1}$ | $\dot{M}_{in}$ 0.01$M_\odot$/yr | $\alpha_{vir}$ | compex | ref |
|---|---|---|---|---|---|---|---|---|---|---|---|---|---|---|
| I09018-4816 | 09:03:32.84 | -48:28:10.0 | 5.2e4 | 1.0e3 | 3.9e5 | 0.22 | 10.38(0.02) | 3.49(0.04) | -0.61 | 3.41(0.13) | 4.75(0.18) | 0.56 | B | 1 |
| I10365-5803 | 10:38:32.46 | -58:19:05.9 | 2.0e4 | 5.0e2 | 1.9e5 | 0.22 | -19.26(0.02) | 2.94(0.04) | 0.83 | -3.30(0.11) | ... | 0.79 | R | 1 |
| I13079-6218 | 13:11:14.16 | -62:34:42.1 | 1.3e5 | 3.1e3 | 1.0e4 | 1.09 | -41.96(0.03) | 5.28(0.06) | 0.19 | ... | ... | 2.01 | N | 2 |
| I13134-6242 | 13:16:43.05 | -62:58:30.1 | 4.0e4 | 1.3e3 | 2.8e4 | 0.57 | -32.21(0.03) | 4.12(0.08) | -0.31 | 1.99(0.09) | 1.35(0.06) | 1.61 | B | 2 |
| I13291-6229 | 13:32:33.30 | -63:45:19.8 | 1.6e5 | 5.0e3 | 1.2e4 | 0.55 | -37.19(0.01) | 2.56(0.03) | 0.06 | ... | ... | 1.50 | N | 2 |
| I14382-6017 | 14:42:02.76 | -60:30:35.1 | 1.6e5 | 4.0e3 | 3.5e3 | 1.68 | -60.17(0.02) | 3.03(0.04) | -0.60 | 2.04(0.06) | 1.49(0.04) | 0.81 | B | 2 |
| I14453-5912 | 14:49:07.77 | -59:24:49.7 | 1.6e4 | 8.0e2 | 6.6e3 | 0.79 | -40.22(0.01) | 2.02(0.02) | 1.16 | -1.05(0.02) | ... | 0.84 | R | 2 |
| I14498-5856 | 14:53:42.81 | -59:08:56.5 | 2.5e5 | 1.0e3 | 1.0e4 | 0.74 | -49.82(0.01) | 3.37(0.02) | -0.66 | 3.77(0.07) | 1.56(0.03) | 1.76 | B | 2 |
| I15520-5234 | 15:55:48.84 | -52:43:06.2 | 1.3e5 | 1.6e3 | 2.2e4 | 0.67 | -41.67(0.01) | 4.53(0.09) | -0.46 | 4.33(0.10) | 3.14(0.07) | 1.82 | B | 2 |
| I15596-5301 | 16:03:32.29 | -53:09:28.1 | 3.2e5 | 8.0e3 | 5.5e3 | 1.81 | -74.20(0.01) | 3.87(0.03) | -0.61 | 3.33(0.05) | 4.48(0.06) | 0.71 | B | 2 |
| I16060-5146 | 16:09:52.85 | -51:54:54.7 | 6.3e5 | 8.0e3 | 1.7e4 | 1.24 | -90.44(0.01) | 5.02(0.04) | -0.56 | 3.91(0.07) | 7.67(0.14) | 0.83 | B | 2 |
| I16071-5142 | 16:11:00.01 | -51:50:21.6 | 6.3e4 | 5.0e3 | 1.2e4 | 1.21 | -86.80(0.02) | 5.80(0.05) | -0.47 | 4.59(0.12) | 5.83(0.15) | 1.70 | B | 2 |
| I16076-5134 | 16:11:27.12 | -51:41:56.9 | 2.0e5 | 4.0e3 | 4.2e3 | 1.57 | -88.37(0.02) | 4.89(0.05) | -0.75 | 6.90(0.20) | 5.37(0.16) | 1.98 | B | 2 |
| I16272-4837 | 16:30:59.08 | -48:43:53.3 | 2.0e4 | 1.6e3 | 1.1e4 | 0.84 | -46.57(0.03) | 3.66(0.07) | -0.40 | 2.58(0.04) | 1.49(0.02) | 1.49 | B | 2 |
| I16344-4658 | 16:38:09.38 | -47:04:58.7 | 2.5e5 | 1.3e4 | 2.6e3 | 2.70 | -49.13(0.03) | 5.04(0.06) | 0.64 | .. | ... | 1.14 | N | 2 |
| I16348-4654 | 16:38:29.64 | -47:00:41.1 | 2.5e5 | 2.5e4 | 7.5e3 | 2.4 | -46.94(0.03) | 5.92(0.06) | -0.51 | 1.81(0.04) | 5.81(0.12) | 0.70 | B | 2 |
| I16351-4722 | 16:38:50.98 | -47:27:57.8 | 7.9e4 | 1.6e3 | 2.0e4 | 0.69 | -40.21(0.01) | 4.59(0.03) | -0.76 | 3.02(0.10) | 2.12(0.06) | 1.93 | B | 2 |
| I16385-4619 | 16:42:14.04 | -46:25:25.9 | 1.3e5 | 1.6e3 | 2.7e3 | 1.34 | -116.97(0.01) | 4.76(0.03) | 0.21 | -2.36(0.08) | ... | 4.02 | R | 2 |
| I16547-4247 | 16:58:17.26 | -42:52:04.5 | 6.3e5 | 1.6e3 | 2.0e4 | 0.69 | -30.36(0.01) | 5.10(0.03) | 1.12 | -4.51(0.07) | ... | 2.37 | R | 2 |
| I17143-3700 | 17:17:45.65 | -37:03:11.8 | 4.0e5 | 6.3e3 | 1.0e3 | 2.95 | -31.83(0.02) | 3.40(0.04) | -0.16 | ... | ... | 1.13 | N | 2 |
| I17160-3707 | 17:19:26.81 | -37:11:01.4 | 1.0e6 | 1.3e4 | 1.1e4 | 1.69 | -70.02(0.02) | 3.84(0.05) | -0.47 | 1.66(0.03) | 3.79(0.06) | 0.42 | B | 2 |
| I17204-3636 | 17:23:50.32 | -36:38:58.1 | 1.6e4 | 7.9e2 | 1.5e4 | 0.60 | -17.74(0.01) | 3.11(0.03) | -0.83 | 1.21(0.03) | 0.49(0.01) | 1.54 | B | 2 |
| I17220-3609 | 17:25:24.99 | -36:12:45.1 | 5.0e5 | 2.0e4 | 5.9e3 | 2.41 | -94.94(0.01) | 4.53(0.03) | -0.77 | 4.71(0.06) | 11.94(0.15) | 0.52 | B | 2 |
| I17271-3439 | 17:30:26.21 | -34:41:48.9 | 4.0e5 | 1.0e4 | 1.7e4 | 1.34 | -15.61(0.01) | 4.20(0.02) | 0.64 | -3.10(0.05) | ... | 0.50 | R | 2 |
| I18056-1952 | 18:08:38.18 | -19:51:49.0 | 5.0e5 | 2.5e4 | 8.3e3 | 2.32 | 68.46(0.07) | 9.07(0.18) | -0.59 | 3.08(0.06) | 10.22(0.19) | 1.59 | B | 2 |
| I18089-1732 | 18:11:51.28 | -17:31:33.4 | 2.0e4 | 1.3e3 | 7.1e3 | 0.90 | 33.42(0.01) | 3.26(0.02) | 1.04 | ... | ... | 0.59 | N | 2 |
| I18116-1646 | 18:14:35.76 | -16:45:40.8 | 1.3e5 | 1.3e3 | 5.3e3 | 0.99 | 48.47(0.01) | 2.46(0.01) | 1.83 | -3.02(0.07) | ... | 1.00 | R | 2 |
| I18182-1433 | 18:21:09.22 | -14:31:46.8 | 2.0e4 | 1.3e3 | 1.4e4 | 0.71 | 59.04(0.02) | 3.41(0.04) | -0.35 | 1.99(0.05) | 1.08(0.03) | 1.37 | B | 2 |
| I18507+0110 | 18:53:18.42 | 01:15:00.1 | 6.3e5 | 1.6e3 | 7.6e4 | 0.44 | 58.50(0.01) | 5.11(0.03) | -0.66 | 4.83(0.04) | 5.33(0.05) | 1.52 | B | 2 |
| I19078+0901 | 19:10:13.41 | 09:06:10.4 | 8.0e6 | 1.0e5 | 5.3e3 | 4.26 | 6.20(0.02) | 13.21(0.05) | -0.60 | .. | ... | 1.56 | R | 2 |

Tablenote: references 1: Urquhart et al. (2018) and 2:Faúndez et al. (2004). $n_{H_2}$ are calculated by use of $n_{H_2} = \frac{M}{4/3\pi R^3 \mu m_H}$. Negative $V_{in}$ indicate outward gas flow or gas expansion. Compex: B denotes "blue profile", R means "red profile" and N indicates the spectra without "blue profile" and "red profile" signatures.

### 3.1 Line profiles

Figure 1 and Figure A.1 present the observed C$^{17}$O(3-2) and CO(4-3) spectra of the 30 sources. The black and green lines are the data of CO(4-3) and C$^{17}$O(3-2) lines, respectively. In total, 24 sources have double-peaked profile in CO (4-3) in which the optically thin C$^{17}$O(3-2) lines peaks near the dips of the CO (4-3) lines. 18 of 24 sources show the "blue profile" (see Figure 1) with the blue peak being stronger than the red one while 6 sources show "red profile" (see Figure A.1 in appendix). The detection rate of "blue profile" is 60%. The other six sources show single or multiple-peaked profile. The observed "blue profile" lines are indicative of gas infall in these massive clumps. The "red profile" lines may indicate gas expansion (Qin

et al. 2016). Some infall sources (I15520, I16071, I16076, I16272, I16348, I17220, I17204) and the other sources (I10365, I16385, I16547, I17271, I16344, I18089, I19078) have strong wing radiations, which may indicate the presence of outflows in these sources. Gaussian fitting was made to the $C^{17}O(3-2)$. The peak velocities and the FWHM line widths were summarized in Table 1. Note that we take peak velocities of the $C^{17}O(3-2)$ as systemic velocities of the clouds.

Mardones et al. (1997) suggested line asymmetry $\delta v$ as an alternative measurement of gas infall for these blue-skewed lines. The line asymmetric parameters are derived by use of the following formula:

$$\delta v = (v_{CO} - v_{C^{17}O})/\Delta v_{C^{17}O}, \quad (1)$$

where $v_{CO}$ and $v_{C^{17}O}$ are the peak velocities of the CO(4-3) and $C^{17}O(3-2)$ lines, respectively, and $\Delta v_{C^{17}O}$ is the FWHM of the $C^{17}O(3-2)$. The resulting line asymmetric parameters are derived and summarized in Table 1.

To express the relative importance of the clump's kinetic and gravitational energy in terms of observable quantities, Bertoldi & McKee (1992) defined the dimensionless virial parameter:

$$\alpha_{vir} = \frac{5\sigma_v^2 R}{GM} \quad (2)$$

Optically thin $C^{17}O(3-2)$ suffurs less from depletion for relatively evolved sources when compared with the clumps at an early stage of star formation. The $C^{17}O(3-2)$ and dust continuum are used for calculating virial parameter of massive clumps before (Giannetti et al. 2014). Here we use $C^{17}O$ for virial parameter calculation and $\sigma_v = \frac{\Delta V_{C^{17}O}}{\sqrt{8\ln 2}}$. The observed $C^{17}O(3-2)$ lines peak around the CO(4-3) absorption dips or the two transitions have same peak velocity for single peak lines, suggesting that $C^{17}O(3-2)$ and CO(4-3) trace same region. $C^{17}O(3-2)$ emission is optically thin, which has similar size with continuum. Giannetti et al. (2014) used $C^{17}O(3-2)$ line to derive $\Delta V_{thin}$. In the calculation, the size (R) used for discussing the virial parameter is taken from the 1.1 mm continuum size, and the mass (M) is also taken from the continuum observations. The calculated values of $\alpha_{vir}$ are showed in Table 1. The source with virial parameter less than 2 is generally taken as indicating that a particular clump is likely to be globally collapsing. From Table 1, one can see that the "blue profile" lines have $\delta v$ less than –0.25 and $\alpha_{vir}$ less than 2, confirming that the observed "blue profile" lines in this work are caused by infall and global collapse is ongoing in these massive clumps. It is interesting to note that all the observed source having $\alpha_{vir}$ less than 2 except for three sources ($\alpha_{vir}$ = 2.01, 2.37, 4.02.) None of the three sources show a blue profile. Outflows as indicated by CO line wings may shape the line profile of the some infall sources in our observations.

### 3.2 Two-layer model

We used two-layer model (Myers et al. 1996; Di Francesco et al. 2001) to fit the observed CO(4-3) "blue profile" and "red profile" spectra to obtain the information of infall and expansion. The model fits the observed spectrum by setting the exciton temperature $J_c$ of the continuum source, the cosmic background radiation temperature $J_b$, the "front" layer excitation temperature $J_f$ and the "rear" layer excitation temperature $J_r$, the beam filling factor $\Phi$, the systemic velocity $V_{LSR}$, the infall velocity $V_{in}$, the velocity dispersion $\sigma$, and the peak optical depth $\tau_0$. The predicted line brightness temperature $\Delta T_B$ at velocity V is given by

$$\Delta T_{\rm B} = (J_f - J_{cr})[1 - e^{(-\tau_f)}] + (1-\Phi)(J_r - J_b)[1 - e^{(-\tau_r - \tau_f)}] \qquad (3)$$

where

$$J_{cr} = \Phi J_c + (1 - \Phi)J_r \qquad (4)$$

$$\tau_f = \tau_0 e^{\left[\frac{-(V-V_{in}-V_{LSR})^2}{2\sigma^2}\right]} \qquad (5)$$

$$\tau_r = \tau_0 e^{\left[\frac{-(V+V_{in}-V_{LSR})^2}{2\sigma^2}\right]} \qquad (6)$$

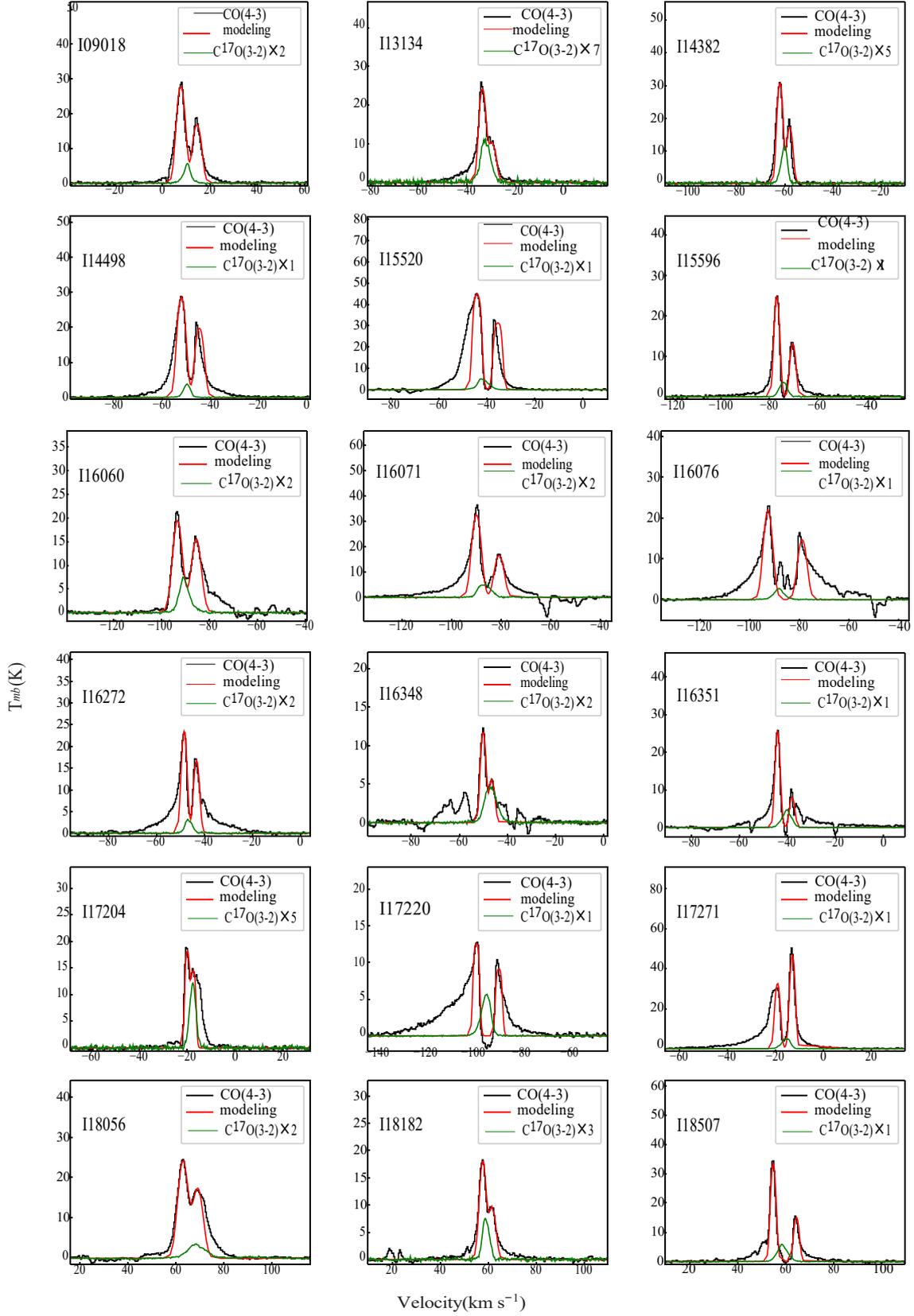

Fig. 1: The observed and modeled spectra for "blue profile" sources. Black and green curves denote the CO (4-3) and C$^{17}$O (3-2) lines. Red curves indicate two-layer model fitting to the CO (4-3) lines.

$J_f = \frac{T_0}{exp(T_0/T)-1}$, where $T_0 = \frac{h\nu}{k}$, h is Planck's constant, k is Boltzmann's constant, and $\nu$ is the transition frequency of CO(4-3). $V_{LSR}$, $\sigma$ and $V_{in}$ determine the line profile, while other parameters determine the intensity of two peaks. In this model, we set the ranges for $\tau_0$ from 0.1 to 4, for the filling factor $\Phi$ from 0 to 1, $J_c$ from 20 to 150 K, $J_b$ bigger than 2.73 K and less than 100 K, $J_r$ from 10 to 80K, $\sigma$ from 0.5 to 3 km s$^{-1}$, $V_{in}$ from 0.5 to 7 km s$^{-1}$, $V_{LSR}$ is fine-tuned by fitting according the value of the peak velocity of C$^{17}$O.

The modelled "blue profile" and "red profile" spectra coded in red color are shown in Figure 1 and Figure A.1, respectively. The infall velocities were determined at a range of 1.21 to 6.90 km s$^{-1}$ and shown in Table 1. Then the mass infall rate $\dot{M}_{in}$ can be calculated by the formula:

$$\dot{M}_{in} = 4\pi R^2 m_H n_{H_2} \mu V_{in} \qquad (7)$$

where $\mu$ =2.8 is weight of hydrogen molecule (Wang et al. 2014), $m_H$ is mass of hydrogen atom. $n_{H_2}$ is the density of hydrogen molecule and R is the source size taken from Table 1. The estimated $\dot{M}_{in}$ values range from $10^{-3}$ to $10^{-1}$ M$_\odot$yr$^{-1}$ which are far larger than those in low-mass star formation regions (Offner et al. 2010; Padoan et al. 2014; Stutz et al. 2013; Beuther et al. 2012). Values of $\dot{M}_{in}$ are listed in Table 1.

## 4 DISCUSSION

The HCN (4-3) lines observed by the ASTE suggested that the 30 sources have blue profile, while only 18 out of the 30 targets show infall signature in CO (4-3). Multiple HCN line observations toward the Sgr B2(M) cloud suggested that global collapse occurs in the colder outer regions, but expansion dominates in the warmer, inner regions (Rolffs et al. 2010). Probably lower detection rate in CO (4-3) line is caused by higher spatial resolution of 13″ by the APEX observations when compared to the HCN (4-3) at an angular resolution of 22″ by the (ASTE) observations (Liu et al. 2016). The lower resolution observations may sample overall gas collapse in outer envelope of the clumps, while higher spatial resolution observations trace inner part of the clouds in which some clumps are expanding.

From Table 1, the clumps with larger mass and higher luminosity tend to have larger mass infall rates.

We used power-law fitting to determine the relationships of the mass infall rate versus the mass and luminosity. Left panel of Figure 2 shows the mass infall rate as a function of the gas mass. The mass infall rate is positively correlated with the gas mass, suggesting that higher mass clumps have higher mass infall rates. Gravitational potential is proportional to the M/R. Righ panel of Figure 2 presents the mass infall rate as a function of M/R, showing similar trend with that in left panel of Figure 2, which may indicate that the gravity plays an important role in the collapsing process in the massive clumps. The higher mass clumps have larger gravitational potential and then have higher mass infall rates. This is consistence with the expectation that the more gravity dominates(Wyrowski et al. 2016).

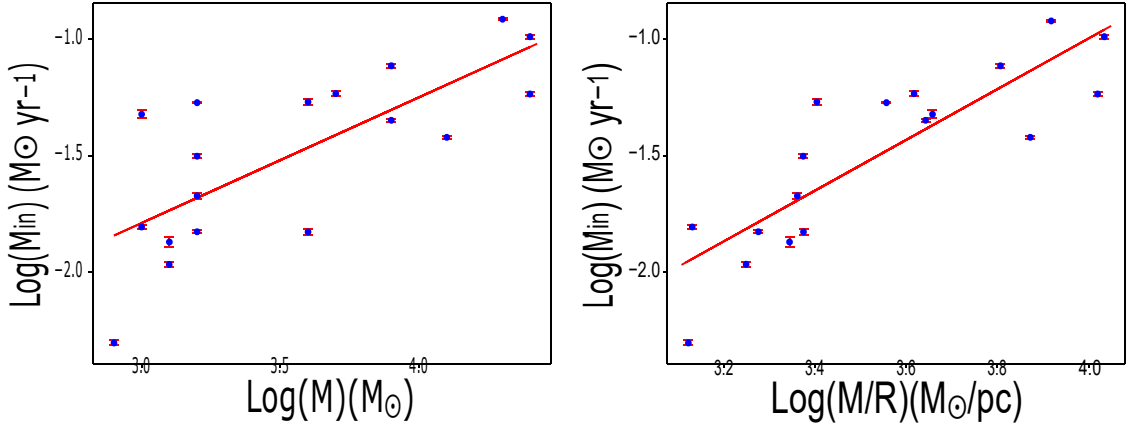

Fig. 2: Left panel: the mass infall rate as a function of the clumps mass. The blue dots with 1 $\sigma$ error bar represent the mass infall rates. The black line shows the power-law fiiting and the Spearman correlation coefficients is 0.55. Right panel: the mass infall rate as a function of M/R . The blue dots with 1 $\sigma$ error bar represent the mass infall rates. The black line shows the power-law fiiting and the Spearman correlation coefficients is 0.73.

The large bolometric luminosity implys that these clumps contain one massive star or several massive stars. Bolometric luminosity is an useful tracer of star formation rate (Kennicutt & Evans 2012). Figure 3 presents a positive trend between mass infall rate and bolometric luminosity. Larger luminosity clumps have higher mass infall rates, suggesting that higher mass infall rate can lead to formation of high-mass stars.

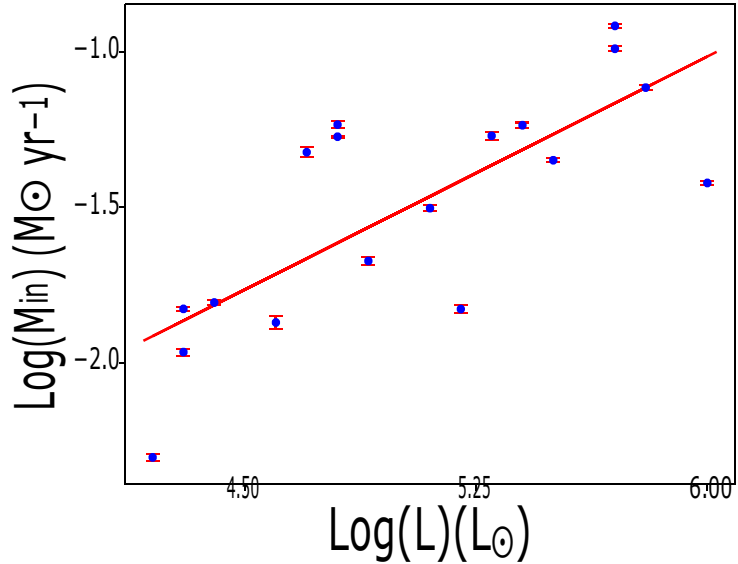

Fig. 3: The mass infall rate as a function of the clump luminosity. The blue dots with 1 $\sigma$ represent the masse infall rates. The black line shows the power-law fitting, and the Spearman correlation coefficients is 0.57.

The observations have shown that mass infall rates did not have obvious variation with evolutionary stage across pre-stellar cores, proto-stellar cores and UC Hii region stages (He et al. 2015). Molinari et al. (2016) suggested that the ratio between the luminosity and the mass of the clump can be used to diagnose the evolutionary stage of massive clumps. There is no correlation between luminosity-mass ratio and mass infall rate in this work as shown in Figure 4 .Similar conclusion is reached by Wyrowski et al. (2016).

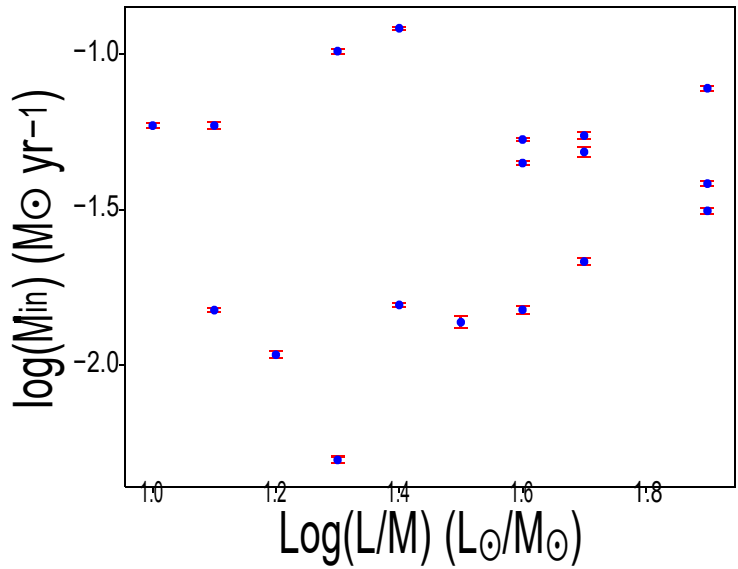

Fig. 4: The mass infall rate as a function of the luminosity-mass ratio. The blue dots with 1 error bar represent the mass infall rates.

## 5 SUMMARY

We have carried out APEX observations in CO(4-3) and $C^{17}O$(3-2) lines to 30 massive clumps to search for infall signatures and investigate infall properties. 18 clumps have the CO(4-3) lines showing blue profile, the line asymmetric parameters less than –0.25 and virial parameter less than 2, which are indicative of golbal collapse in these massive clumps. Larger mass infall rates are derived. The clumps with higher gas mass and luminosity have larger mass infall rate, indicating that the gravity plays a dominant role in the collapsing process and higher mass infall rate can lead to formation of high-mass stars. The values of mass infall rate are not related to evolutionary stage, implying that gas infall can occur at various evolutionary stage.

**Acknowledgements** This work has been supported by the National Key Reasearch and Development Program of China (No. 2017YFA0402701, 2019YFA0405100) and by the National Natural Science Foundation of China (U1631237, 11973013, 11721303 11373026 and 11433004), and the starting grant at the Kavli Institute for Astronomy and Astrophysics, Peking University (7101502016).

**Appendix A: APPENDIX**

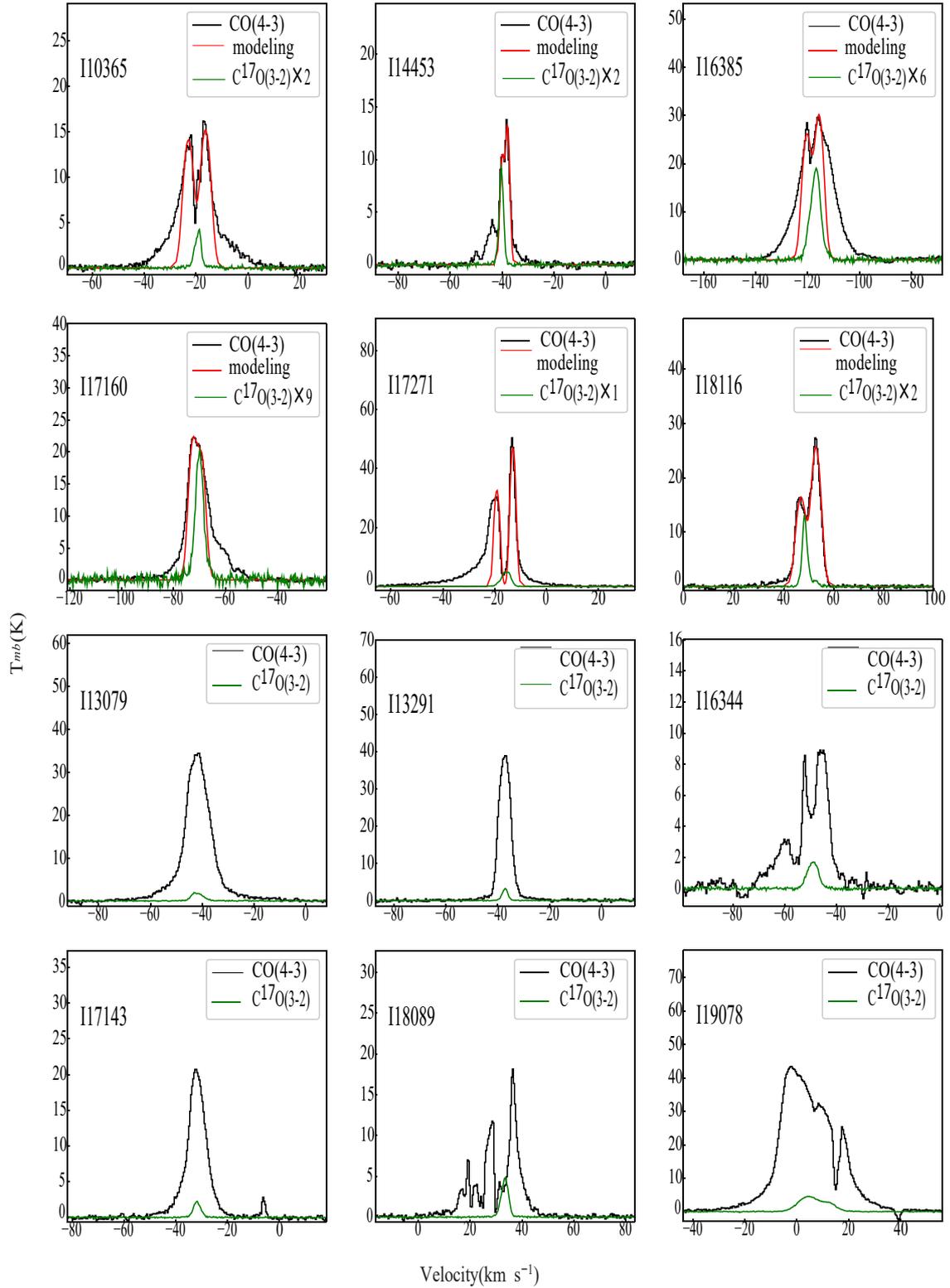

Fig.A.1: The observed and modeled spectra for "red profile" sources. Black and green curves denote the CO (4-3) and $C^{17}O$ (3-2) lines. Red curves indicate two-layer model fitting to the CO (4-3) lines.